\documentclass[aps,twocolumn,floatfix,prb,superscriptaddress]{revtex4}
\usepackage{epsfig}

\begin{document}

\title{Grain boundaries in vortex matter}

\author{Paolo Moretti}
\affiliation{Center for Materials Science and Engineering,
University of Edinburgh,
King's Buildings, Sanderson Building, Edinburgh EH93JL, UK}
\author{M.-Carmen Miguel}
\affiliation{Departament de F\'{\i}sica Fonamental,
Facultat de F\'{\i}sica, Universitat de Barcelona, Av. Diagonal 647,
E-08028, Barcelona, Spain}
\author{Stefano Zapperi}
\affiliation{INFM UdR Roma 1 and SMC, Dipartimento di Fisica,
Universit\`a "La Sapienza", P.le A. Moro 2, 00185 Roma, Italy}
\affiliation{Istituto dei Sistemi Complessi, CNR, via dei Taurini
19, 00185 Roma, Italy}
\begin{abstract}
We explore the statistical properties of grain boundaries in the
vortex polycrystalline phase of type II superconductors. Treating
grain boundaries as arrays of dislocations interacting through linear
elasticity, we show that self-interaction of a deformed grain boundary
is equivalent to a non-local long-range surface tension. This affects
the pinning properties of grain boundaries, that are found to be less
rough than isolated dislocations. The presence of grain boundaries has
an important effect on the transport properties of type II
superconductors as we show by numerical simulations: our results
indicate that the critical current is higher for a vortex polycrystal
than for a regular vortex lattice. Finally, we discuss the possible
role of grain boundaries in vortex lattice melting. Through a
phenomenological theory we show that melting can be preceded by an
intermediate polycrystalline phase.
\end{abstract}
\maketitle

\section{Introduction}

The phase diagram of high temperature superconductors is an object of
active investigation in condensed matter physics. Most high $T_c$
materials behave in a magnetic field as type II superconductors, with
further complications due to the broader phase space --- in terms of
temperature $T$ and field $H$ --- in comparison to conventional type
II superconductors \cite{BLA-94,BRA-95,GIA-01}. This leads to several
possibilities for the mixed phase, where magnetic flux penetration is
incomplete. As first discussed by Abrikosov for conventional
superconductors \cite{ABR-57}, flux is quantized and carried by vortex
lines which are arranged in the form of a lattice. As in conventional
matter strong enough fluctuations destroy long range order: when
temperature is raised the vortex lattice melts into a vortex liquid
\cite{SAF-92,BOC-01,AVR-01}. Fluctuations are also provided by
quenched disorder that is intrinsically present in these materials,
leading to complex glassy phases
\cite{CUB-93,GAM-98,LIN-01,GIA-94,FIS-91,NEL-00}.

While several experimental methods have been used to investigate
vortex matter, a direct image of the geometrical and topological
properties of the vortices can be obtained by the Bitter decoration
technique \cite{GRI-94}. Its application to conventional
superconductors provided the first direct proof of the vortex lattice
\cite{ESS-67} predicted by Abrikosov \cite{ABR-57}. The observed
lattice contains, however, topological defects, such as dislocations
and grain boundaries. These last extended defects are the signature of
a vortex polycrystal with crystalline grains of different orientations
\cite{GRI-94,GRI-89}.  Vortex polycrystals have been observed, after
field cooling, in various superconducting materials such as NbMo
\cite{GRI-94,GRI-89}, NbSe$_2$ \cite{MAR-97,MAR-98,PAR-97,FAS-02},
BSCCO \cite{DAI-94} and YBCO \cite{HER-00}. The grain size is
typically found to grow with applied magnetic field
\cite{MAR-97,GRI-89}. Moreover, two-sided decoration experiments show
that the grain boundaries thread the sample from top to bottom
\cite{MAR-97,MAR-98}, i.e., one observes a columnar grain structure.
Despite the wealth of experimental observations, there is no detailed
theory accounting for the formation of vortex polycrystals.

The behavior of vortex matter in presence of disorder represents a
formidable theoretical problem that has still not been completely
solved. While early theoretical considerations seemed to imply that
even a small amount of disorder would lead to the loss of long-range
order \cite{LAR-70} and to the formation of an amorphous vortex glass
phase \cite{FIS-91}, it is now accepted that at low disorder vortices
arrange into a topologically ordered phase: the Bragg glass
\cite{GIA-94,GIA-95}. The existence of this phase, characterized by
logarithmically growing correlations, slow relaxation, and other
glassy features, has been now experimentally confirmed
\cite{KLE-01}. At high enough disorder, the Bragg glass phase is found
to be unstable against dislocation proliferation and one may expect
the transition into an amorphous vortex glass
\cite{CAR-96,KIE-97,FIS-97}. The precise nature of this transition
and, more generally, the mechanism underlying vortex lattice melting
is still under debate. Typical melting theories are based on variants
of the Lindemann criterion with disorder \cite{MIK-03}, or involve
dislocation proliferation mechanisms \cite{KIE-00}.

The properties of dislocations in the vortex lattice have been the
object of extensive theoretical investigations
\cite{NAB-80,BRA-86,MIG-97,KIE-00b}, but grain boundaries are less
studied although they are often observed in numerical simulations
\cite{MIG-04,DAS-04,CHA-04}.  For instance, the vortex plastic flow in
the Corbino disk geometry is characterized by radial grain boundaries
sliding in the tangential direction \cite{MIG-04}. In addition, recent
numerical simulations indicate the presence of an intermediate
polycrystalline phase before the melting transition
\cite{DAS-04,CHA-04}. This behavior was observed using different
numerical methods in two dimensions \cite{CHA-04} and in presence of
columnar disorder \cite{DAS-04}. This suggests that, in some
conditions, grain boundaries may play a role in the melting process,
as in the theory of grain boundary induced melting of two dimensional
crystals \cite{CHU-83}.

Here we analyze the properties of grain boundaries in vortex matter
describing the fluctuations induced by disorder, stress or
temperature. A grain boundary can be considered as an array of
dislocations, whose dynamics is ruled by internal stresses. While
ideally a grain boundary minimizes its energy by remaining flat, the
action of external perturbations leads to deformations that can be
described by the theory of elasticity \cite{MOR-03}.  We compute the
self-interaction of a deformed grain boundary extending the results
obtained for isotropic elasticity \cite{MOR-03} to the case of the
vortex lattice.  Grain boundaries are much stiffer than isolated
dislocations, possessing a non-local long-range surface tension and,
in presence of disorder, they are expected to be less rough than
isolated dislocations. We estimate the grain boundary roughness
exponent using the random stress model introduced in
Ref.~\onlinecite{KIE-00b} for vortex dislocations. Using scaling
arguments, we also derive the creep law for thermal activated motion
and discuss disorder arrested grain growth \cite{MOR-04}.

The critical current is an important property of type II
superconductors, since it represents the current below which
vortices are pinned and the material conducts without resistance.
It is thus interesting to understand how the topological
properties of vortex matter influence its behavior.  We use
numerical simulations of interacting vortices to quantify the
effect of grain boundaries on the critical current. We obtain a
polycrystalline vortex structure by relaxing at zero temperature a
random initial vortex arrangement. This process simulates a
typical field cooling experiment in which the temperature is
rapidly decreased from above $T_c$ in presence of a field. The
system moves rapidly towards lower energy configurations
corresponding to zero temperature and thermal effects can thus be
disregarded. In this case magnetic flux is present in the material
as it enters the superconducting phase and vortices are initially
disordered. Once grain growth has stopped, we simulate the effect
of an external current flowing through the sample by applying a
constant Lorenz force. The critical current is then defined as the
current at which vortices start to move steadily. By repeating the
simulations for different values of the vortex number,
representing the effect of various magnetic field intensities, we
show that the critical current for a polycrystal is always larger
than the one obtained for a perfect lattice. This reflects the
fact that a polycrystalline assembly is more effectively pinned
than a perfect lattice because it can accommodate better in the
disordered landscape. In addition, we find that the corresponding
IV curve is hysteretic upon ramping up and down the current. This
result can explain the difference in transport properties between
field cooled and zero field cooled samples and the related
hysteresis commonly measured experimentally
\cite{XIA-01,HEN-96,XIA-00,BAN-99,JOU-00,LIN-01,SAS-00}.

Finally, we discuss the possible role of grain boundary in vortex
lattice melting by constructing a free energy functional for the
grain boundary density along the lines of
Ref.~\onlinecite{CHU-83}. We derive the contribution due to grain
boundary fluctuations and junction formation and show the presence
of a polycrystalline phase with a finite grain boundary density.
As temperature is increased the grain boundary density increases
and the system melts.  The theory predicts a value for the melting
temperature that is quite similar to that obtained in
Ref.~\onlinecite{KIE-00} considering only the contribution of
isolated dislocations. We notice that the presence of an
intermediate polycrystalline phase in vortex lattice melting was
recently proposed on a phenomenological basis in
Ref.~\cite{MEN-01} and experimental evidence was reported for
La$_{1.9}$Sr$_{0.1}$CuO$_4$ \cite{DIV-04}.

The paper is organized as follows: in section \ref{elasticity} we
compute the self-energy of a deformed vortex grain boundary.  In
section \ref{disorder_section} we analyze the interaction between
grain boundaries and disorder, as well as its relevance for pinning,
creep, and grain growth. Section \ref{simulations} reports the results
of numerical simulations of interacting vortices where we discuss the
effect of grain boundaries on the critical current. In section
\ref{melting} we discuss the role of grain boundaries in the melting
process through a phenomenological theory. Section \ref{conclusions}
is devoted to conclusions.  Finally, the appendix reports details of
the derivation reported in section \ref{melting}.

\section{Elasticity of grain boundaries}\label{elasticity}

A simplified but rather effective description of the vortex lattice is
provided by its representation as an elastic crystal of flux lines. At
large enough distances, the elastic energy of the vortex lattice can
be expressed in terms of the vortex displacement field ${\bf u}$ as
follows
\begin{equation}\label{H}
\mathcal{H}=\frac{1}{2}\int d^3r\left[c_{66}({\bf\nabla
u})^2+(c_{11}-c_{66})({\bf \nabla\cdot u})^2+
c_{44}(\partial_z{\bf u})^2\right],
\end{equation}
where $c_{11}$, $c_{44}$, $c_{66}$ are the local elastic moduli, and
the magnetic induction ${\bf B}$ is parallel to the $z$
direction. Within this representation, we shall introduce an ideal low
angle grain boundary as an infinite periodic array of straight
dislocations in the vortex lattice oriented along the $z$ axis,
spatially arranged along the $y$ axis with an array spacing equal to
$D$, and with Burgers vectors ${\bf b}$ pointing along the $x$
direction (i.e. edge dislocations). The wandering of the $i$-th
dislocation line can be schematized through the vector ${\bf
R}_i(z)=(X_i+X_i(z),iD)$, assuming that all displacements take place
within glide planes, i.e. the $xz$ plane, so that $X_i+X_i(z)$ plays
the role of the displacement field of the grain boundary as well.
$X_i$ is a constant term and deals with rigid displacements of the
dislocation lines. Its contribution to the elastic Hamiltonian is
known since it is the same as for straight dislocations in isotropic
lattices\cite{MOR-03}. In the following this contribution will be
referred to as $\mathcal{H}_0$.

Defining ${\bf r}_\perp=(x,y)$, the vector ${\bf u}$ can be
decomposed as ${\bf u}({\bf r})={\bf u^r}({\bf
r_{\perp}},z)+\sum_i{\bf u_i^s}({\bf r_{\perp}}-{\bf R_i}(z),z)$,
where ${\bf u_i^s}({\bf r_{\perp}}-{\bf R_i}(z),z)$ is the singular
solution of the two-dimensional problem for each value of $z$
\begin{eqnarray}\label{p2d} \left\{\begin{array}{cc}
c_{66}{\bf\nabla}^2{\bf u}_i^s +(c_{11}-c_{66}){\bf\nabla}
({\bf\nabla \cdot u}_i^s)=0\\
\oint d{\bf u}_i^s ={\bf b}_i \hspace{1.5cm}\forall i,z
\end{array}\right.
\end{eqnarray}
while ${\bf u^r}({\bf r_{\perp}},z)$ is the regular part of the solution
due to the interplane couplings along $z$.

Minimizing Eq.(\ref{H}) with respect to ${\bf u}$ and imposing the
first expression of Eqs.~(\ref{p2d}) we find the differential equation
\begin{equation}\label{diff}
c_{66}{\bf\nabla}^2{\bf u}^r+(c_{11}-c_{66}){\bf\nabla}({\bf\nabla\cdot
u}^r)+c_{44}\partial^2_z{\bf u}^r =-c_{44}\partial^2_z\sum_i{\bf
u}_i^s,
\end{equation}
where the field ${\bf u}_i^s$ on the right-hand side term of the
equation is known from elasticity theory as the displacement field
generated by a point edge dislocation at ${\bf R_i}(z)$. Performing a
first order expansion in the displacement $X_i+X_i(z)$, the derivative
removes any dependence on the constant part $X_i$, and we can rewrite
Eq.(\ref{diff}) in Fourier space as follows
\begin{equation}
c_{66}q^2{\bf u}^r+(c_{11}-c_{66}){\bf q}({\bf q\cdot
u}^r)+c_{44}k_z^2{\bf u}^r= c_{44}\frac{k_z^2}{q^2}{\bf A},
\end{equation}
where ${\bf q}=(k_x,k_y)$ and
\begin{eqnarray}
{\bf A}=\sum_ne^{ik_yY_n}X_n(k_z)\left(\begin{array}{cc}
k_y[r-(1-r)\cos2\phi]\\k_x[r+(1-r)\cos2\phi]\end{array}\right)
\end{eqnarray}
with $r=c_{66}/c_{11}$, $\cos\phi=k_x/k$ and $\sin\phi=k_y/k$.

${\bf A}$ can be decomposed in its longitudinal and transverse
components ${\bf A}_L={\bf q}({\bf q}\cdot{\bf A})/q^2$ and ${\bf
A}_T={\bf A}-{\bf A}_L$. The Hamiltonian (\ref{H}) thus becomes
\begin{widetext}\begin{equation}
\mathcal{H}=\mathcal{H}^0+\frac{1}{2}\,c_{44}\,b^2\sum_{n,m}\int
\frac{d^2q}{(2\pi)^2}\int\frac{dk_z}{2\pi} \,k_z^2
\,M(q,\phi,k_z)\,e^{ik_y(n-m)D}\,X_n(k_z)X_m(-k_z)
\end{equation}
where we have neglected constant terms and defined
\begin{equation}
M(q,\phi,k_z)\equiv\left[
\frac{c_{66}\cos^22\phi}{c_{66}q^2+c_{44}k_z^2}+\frac{c_{11}r^2\sin^22\phi}{c_{11}q^2+c_{44}k_z^2}\right].
\end{equation}
Defining $\displaystyle
X_n(k_z)=\int_{BZ}\frac{dQ_y}{2\pi}\,e^{-iQ_ynD}\,X(Q_y,k_z)$, where the
integral is restricted to the first Brillouin zone (BZ), we get
\begin{equation}
\mathcal{H}=\frac{1}{2}\,c_{44}\,\frac{b^2}{D^2}\sum_{G_y}\int_{BZ}\frac{dQ_y}{2\pi}
\int\frac{dk_z}{2\pi}\,\Xi(Q_y+G_y,k_z)\,X(Q_y,k_z)X(-Q_y,-k_z)
\end{equation}
where we have introduced the interaction kernel
\begin{equation}\label{integral}
\Xi(Q_y+G_y,k_z)=k_z^2\int_{-\infty}^{+\infty}M(k_x,Q_y+G_y,k_z)\,dk_x
\end{equation}
with
\begin{equation}
M(k_x,k_y,k_z)=\frac{(k_x^2-k_y^2)^2}{(k_x^2+k_y^2)^2(k_x^2+k_y^2+\frac{c_{44}}{c_{66}}k_z^2)}+4r^2
\frac{k_x^2k_y^2}{(k_x^2+k_y^2)^2(k_x^2+k_y^2+\frac{c_{44}}{c_{11}}k_z^2)}.
\end{equation}
Solving the integral in Eq.(\ref{integral}) leads to
\begin{eqnarray}\label{leading}
\mathcal{H}=\mathcal{H}^0+\frac{\pi b^2}{2D^2}\,\frac{c_{66}^2}{c_{44}}\,\sum_{G_y}\int_{BZ}\frac{dQ_y}{2\pi}
\int\frac{dk_z}{2\pi}\,X(Q_y,k_z)X(-Q_y,-k_z)\nonumber\\
\frac{1}{k_z^2}\left[\frac{\left(2k_y^2+\frac{c_{44}}{c_{66}}k_z^2\right)^2}
{\sqrt{k_y^2+\frac{c_{44}}{c_{66}}k_z^2}}
-4k_y^2\sqrt{k_y^2+\frac{c_{44}}{c_{11}}k_z^2}-2\left(\frac{c_{44}}{c_{66}}
-\frac{c_{44}}{c_{11}}\right)
|k_y|\,k_z^2\right]
\end{eqnarray}
with $k_y=Q_y+G_y$, $\;\displaystyle\frac{c_{44}}{c_{66}}\gg 1\;\;\mbox{and}\;\;\frac{c_{44}}{c_{11}}\sim 1$.

Moreover, keeping the leading term of the righthand side in
Eq.(\ref{leading}) we get

\begin{equation}
\mathcal{H}_{GB}=\frac{\pi b^2}{2D^2}\sum_{G_y}\int\frac{dQ_y}{2\pi}
\int\frac{dk_z}{2\pi}\, (
2c_{66}|k_y|+\sqrt{c_{44}c_{66}}|k_z|)\,X(k_y,k_z)X(-k_y,-k_z).\label{eq:elast_gb}
\end{equation}
\end{widetext}

It is a common procedure to rescale the $y$ coordinate by a factor
$\displaystyle \frac{1}{2}\sqrt{\frac{c_{44}}{c_{66}}}$ \cite{BLA-94},
in order to get an isotropic reference frame. The elastic Hamiltonian
thus becomes
\begin{equation}\label{iso}
\mathcal{H}=K\,\frac{\pi b^2}{2D^2}\sum_{G_y}\int\frac{d^2k}{(2\pi)^2}
\,|{\bf k}|\,X({\bf k})X(-{\bf k}),
\end{equation}
being ${\bf k}=(k_y,k_z)$ and $K=\sqrt{c_{44}c_{66}}$.

In this limit, the same result predicted by the isotropic
theory\cite{MOR-03} is thus obtained. The nonlocal character of the
elastic kernel ($\propto k$) manifests that long range interactions
between dislocations stiffen the grain boundary, and that a surface
tension approximation is not suitable for a correct description of its
elastic properties.

\section{Interaction between grain boundaries and disorder}
\label{disorder_section}

\subsection{Random stresses}

Point defects such as vacancies or interstitials in the underlying
crystalline structure of the superconducting material, and/or
substitutional impurities, etc., act as pinning centers for the
magnetic vortices. For weak pinning forces, disorder can be
theoretically described by a {\em random pinning potential} acting
directly on flux lines. The distortions generated in the vortex
lattice as well as the occurrence of depinning under an applied
current have been intensively studied over the last decades (for a
review see Ref.~\onlinecite{BLA-94}).

Here, we are instead concerned with the behavior of grain boundaries
in presence of disorder. The disorder induced vortex lattice
displacement field gives rise to shear elastic stresses, which, in
turn, generate Peach-Koehler forces on the vortex lattice
dislocations~\cite{HIR}. In other words, as the final consequence of these
disorder-induced distortions, there is an effective pinning stress
field $\sigma_{ij}({\bf r})$ acting as well on vortex dislocations
(and therefore on grain boundaries). The statistical properties of the
random stress field has been analyzed in Ref.~\onlinecite{KIE-00b} in
the case of vortex dislocations. In the following, we recall their
derivation and adapt it to the case of grain boundaries.

On short length scales, where vortex displacements ${\bf u}({\bf r})$
are smaller than the coherence length $\xi$ (the so called Larkin
regime \cite{LAR-79}), a perturbative calculation can be performed. As
discussed in Ref.~\onlinecite{KIE-00}, for grain boundaries it is
necessary to consider larger scales, $\xi <u<a$, where vortices are
well described by a {\it Random Manifold} (RM)
model~\cite{BLA-94,GIA-94} in which flux lines are subject to an
uncorrelated pinning potential. In this case, the relative
displacements correlation function is
\[ B_{ij} ({\bf r}-{\bf r'})=\overline{[u_i({\bf
r})-u_i({\bf r'})][u_j({\bf r})-u_j({\bf r'})]}\]\begin{equation}
\simeq a^2\left(\frac{r-r'}{R_a}\right)^{2\zeta_{RM}}.
\end{equation}
Here $R_a$ is the crossover length, also known as {\it positional
correlation length}, at which average vortex displacements are of the
order of $a$.  The roughness exponent can be estimated as
$\zeta_{RM}\approx 1/5$.

On scales larger than $R_a$, vortex displacements are of the order of
$a$ and the periodicity of the lattice comes into play~\cite{GIA-94}.
Displacements are shown to grow logarithmically, with correlations of
the form $\displaystyle B({\bf r}-{\bf r'})\simeq
\left(\frac{a}{\pi}\right)^2 \,\ln\frac{e|r-r'|}{R_a}$, and
topological defects are absent.  This quasi-ordered phase is knwon as
the {\it Bragg glass} (BrG) \cite{GIA-94}.

The defect-free regions discussed above act on
vortex lattice dislocations through a Peach-Koehler stress
field~\cite{HIR}. Statistical properties of this stress field can be obtained
from the correlator $B_{ij} ({\bf r}-{\bf r'})$ applying linear
elasticity theory.  In particular, the stress correlator
$S_{xy}({\bf r}-{\bf r'})= \overline{\sigma_{xy}({\bf
r})\sigma_{xy}({\bf r'})}$ will read
\begin{eqnarray}
S_{xy}({\bf r}-{\bf r'})=
(K^2/2)\left[\partial_x\partial_{x'}B_{yy}({\bf r}-{\bf r'})+\right.\nonumber\\
\left.\partial_y\partial_{y'}B_{xx}({\bf r}-{\bf r'})+2\partial_x\partial_{y'}B_{yx}
({\bf r}-{\bf r'})\right].
\end{eqnarray}

Replacing previous expressions of $B_{ij}({\bf r}-{\bf r'})$ we easily
obtain the stress fluctuations over a distance $R$
\begin{eqnarray}
S_{xy}(R)\approx K^2\,\frac{a^2}{R^2}\left\{
\begin{array}{cr}
(R/R_a)^{2\zeta_{RM}} & R<R_a\\
1 & R>R_a
 \end{array}
\right.\end{eqnarray} where the first case applies to the RM
description, while the second corresponds to the BrG regime.  The
effect of this random stress on isolated dislocations was studied in
Ref.~\onlinecite{KIE-00b} where several differences with respect to
the case of vortex lines were pointed out. Here we consider the
behavior of grain boundaries, expecting substantial novel features
arising from long range interactions between grain boundary
dislocations.

The Hamiltonian of a grain boundary in presence of disorder can be written as
\begin{equation}
\mathcal{H}_d=\mathcal{H}+\mathcal{H}_{pin},
\end{equation}
with $\mathcal{H}$ being the elastic term calculated above and
$\mathcal{H}_{pin}$ the pinning term given by
\begin{equation}
\mathcal{H}_{pin}=\sum_i\int dz \,X_i(z)\,b\sigma_{xy}[X_i(z),iD,z].
\end{equation}

Although there is no explicit expression for $\mathcal{H}_{pin}$, it
is possible to derive its fluctuations over a distance $L$ as
\begin{widetext}
\begin{equation}
E^2_{pin}=b^2\sum_{i,i'}\int_{0}^{L}\!dz\int_{0}^{L}\! dz'\,X_i(z)X_{i'}(z')\,\overline{\sigma_{xy}[X_i(z),iD,z]
\sigma_{xy}[X_{i'}(z'),i'D,z']}.
\end{equation}
\end{widetext}
Taking the continuum limit of the sum and integrating for both the RM
and the BrG regimes, the typical pinning energy when displacing a
grain boundary segment of length $L$ by an amount $\overline{X_i^2}^{1/2} \sim u_{GB}$ will be
thus given by
\begin{eqnarray}\label{pinning}
E^2_{pin}\simeq \left(\frac{K ab}{D}\right)^2L^2\,u^2_{GB}\left\{
\begin{array}{cc}
\left(\frac{L}{R_a}\right)^{\frac{2}{5}} &\mbox{(RM)}\\\\
\mbox{ln}\,\frac{L}{u_{GB}} & \mbox{(BrG)}
\end{array}\right.
\end{eqnarray}
A dimensional estimate of the elastic cost of
fluctuations of a grain boundary fraction of linear dimension $L$ has
the form
\begin{equation}\label{elastic}
E_{el}=\frac{Kb^2}{D^2}Lu_{GB}^2.
\end{equation}

Since we are dealing with static properties of the system, we can
impose equilibrium conditions balancing $E$ and $E_{pin}$, that is,
equating the elastic cost of fluctuations and the energy gain due to
the interaction with disorder.  Defining the roughness exponent of a
grain boundary $\zeta_{GB}$ from $u_{GB}^2 \sim
L^{2\zeta_{GB}}$ we get
\begin{eqnarray}
\zeta_{GB}\,\approx\,\left\{
\begin{array}{cc}
\frac{1}{5}&\mbox{(RM)}\\\\
\mbox{log}^{1/2}&\mbox{(BrG)}
\end{array}\right.\end{eqnarray}
The long-range stiffness of a grain boundary reduces the values of
roughness exponents in comparison with the case of isolated
dislocations~\cite{KIE-00b}.

\begin{table*}\label{exponents}
\begin{tabular}{| c | c | c | c | c |}
\hline exponent & length scale & isolated dislocation & 2D bundle & grain boundary\\
\hline
\hline $\zeta$ & RM  & $15/13$        & $5/13$  & $1/5$       \\
\hline $\zeta$ & BrG & $1-\log^{2/3}$ & $1/3$   & $\log^{1/2}$\\
\hline $\mu$   & RM  & $17/11$        & $10/21$ & $7/4$       \\
\hline $\mu$   & BrG & $1$            & $2/5$   & $1$         \\
\hline
\end{tabular}
\caption{Comparison between roughness and creep exponents calculated
for isolated dislocations, 2D dislocation bundles~\cite{KIE-00b}, and low
angle grain boundaries, taking into account non-local effects proven
in Section \ref{elasticity}.}
\end{table*}

\subsection{Depinning and creep}

So far we have not considered the effect of  driving forces on the
dislocation arrangement. Driving forces for grain boundary motion can
be externally induced by a current flowing in the superconductor or
internally generated by the ordering process during grain growth
\cite{HAZ-90}. In both cases, the presence of a driving shear stress
$\sigma$ gives rise to a Peach-Koehler force per unit length of the
form $F_{drive}=\sigma b$ acting on each dislocation along the grain
boundary fraction considered or, in other words, to a total driving
force per unit length equal to $F_{drive}=\sigma b L/D$.

At low stress grain boundaries are pinned. One can estimate the
depinning stress from conventional scaling arguments.  The energy
associated to the driving force acting on a low-angle grain boundary
segment of length $L$ and displaced by an amount $u_{GB}$ is given by
\begin{equation}
E_{drive}(L)=\sum_i\int dz\,F_{drive}^i(z)\,u_{GB}(y_i,z)\sim \frac{\sigma b L^2}{D}u_{GB}.
\end{equation}
The depinning stress can be obtained comparing this driving term with
the pinning energy reported in Eq.~\ref{pinning}. The relevant scale
to consider is due to the interplay between elasticity and disorder
and results from the minimization of $E_{el}+E_{pin}$ for
displacements of the order of $u_{GB}\simeq a \simeq b$, corresponding
to the dislocation core. A similar approach is followed in the case of
vortices \cite{LAR-79}, which are pinned for displacements of the
order of $\xi$, the size of the vortex core and, hence, the relevant
scale for the interaction with impurities. In our case, we obtain the
Larkin length as $L_p \simeq (b/D)^5 R_a$, which is typically smaller
than $R_a$.  The depinning stress is then identified as the stress
necessary to depin a section of dimension $L_p$:
\begin{equation}\label{eq:s_c}
\sigma_c\simeq Kb^2/(D L_p)= K D^4/(b^3 R_a).
\end{equation}

For low values of the stress ($\sigma \ll \sigma_c$), the response of
a grain boundary is mainly due to thermally activated motion in a
disordered environment \cite{BLA-94}. In this case, we expect a highly
non-linear creep motion with an average velocity $v\sim
\exp[-C(\sigma_c/\sigma)^{\mu}/T]$, where $C$ is a constant, and $\mu$
is the creep exponent that quantifies the divergence of the energy
barriers $U(\sigma)\sim \sigma^{-\mu}$ separating metastable
states. An estimation of the exponent $\mu$ for a grain boundary can
be obtained from a simple dimensional scaling argument, which is
confirmed by a more rigorous renormalization group analysis.  The
typical energy barrier for a grain boundary section of length $L$ is
of the order of $U(L) \sim L^{1+2\zeta_{GB}}$, where we have used
$u_{GB} \sim L^{\zeta_{GB}}$. In presence of an applied stress
$\sigma$, we can compute the typical grain boundary length $L(\sigma)$
involved in thermal activated motion minimizing
$U(L)+E_{drive}(L)$. The result yields $L(\sigma) \sim
\sigma^{1/(\zeta_{GB}-1)}$.  Using this length, we obtain that the
typical energy barrier depends on the stress as
\begin{equation}
U(\sigma) \sim \sigma^{(1+2\zeta_{GB})/(\zeta_{GB}-1)},
\end{equation}
implying that $\mu=2\zeta_{GB}/(2-\zeta_{GB})$. For the RM and
BrG regimes the exponents are given by
\begin{eqnarray}
\mu_{pl}\approx\left\{\begin{array}{cc}\frac{7}{4}&\mbox{(RM)}\\1&\mbox{(BrG)}\end{array}.\right.
\end{eqnarray}
Now these exponents are larger than their counterparts calculated for
isolated dislocations.  In other words, the formation of grain
boundaries affects vortex dynamics lowering ordinary creep rates.  On
table~\ref{exponents}, all previous results are summarized and
compared to estimates for different dislocation arrays.

\subsection{Grain growth}

In a field cooling experiment, magnetic flux is already present in
the sample as it is quenched in the mixed superconducting phase.
It is thus reasonable to expect that vortices are originally
disordered and that, due to their mutual interactions, undergo a
local ordering process. Along this process, many dislocations
annihilate, and most of the remaining dislocations arrange
themselves into grain boundaries with various orientations. The
growth of crystalline vortex grains is due to the motion of these
separating boundaries. The resulting polycrystalline structure has
been indeed observed experimentally by means of Bitter decorations
of both high~\cite{DAI-94,HER-00} and low
$T_c$~\cite{MAR-97,MAR-98,FAS-02} superconducting samples.  The
effect of quenched disorder is to pin the grain boundaries,
hindering the growth process. Thus to understand the properties of
vortex polycrystals, it is important to analyze the dynamics of
grain boundaries in vortex matter as they interact with
disorder~\cite{MOR-04}.

Grain growth is driven by a reduction in energy: For an average grain
size $R$ and straight grain boundaries, the characteristic energy
stored per unit volume in the form of grain boundary dislocations is
of the order of $\Gamma_0/R$, where $\Gamma_0$ is the energy per unit
area of a grain boundary. Hence, the energy gain achieved by
increasing the grain size by $dR$ is $\Gamma_0/R^2dR$. Physically, the
removal of grain boundary dislocations occurs through the motion of
junction points in the grain boundary network. As junction points must
drag the connecting boundary with them, which may be pinned by
disorder, motion can only occur if the energy gain at least matches
the dissipative work which has to be done against the pinning
forces. The dissipative work per unit volume expended in moving all
grain boundaries by $dR$ is $\sigma_c b/(DR)dR$, where $\sigma_c$ is
the pinning force per unit area. Balancing against the energy gain
yields the limit grain size
\begin{equation}
R_g \approx \frac{D \Gamma_0}{b \sigma_c}\;.\label{eq:rg}
\end{equation}
An explicit expression for the grain size can be obtained inserting
$\sigma_c$, reported in Eq.~\ref{eq:s_c} for the weak pinning regime,
yielding $R_g \propto R_a $. A similar calculation can be performed in
the strong pinning regime \cite{MOR-04} and the results appear to be
in good agreement with experiments measuring average grain sizes in
NbMo \cite{GRI-89}.

\section{The effect of grain boundaries on the critical current}
\label{simulations} Numerical simulations of interacting vortices
in two dimensions (2D) allow to verify and to keep track of the
ordering process and the grain formation after a rapid field
cooling of the vortex system in thin superconducting films. To
this end, we consider a square 2D superconducting cross-section of
linear dimension $L$ perpendicular to the external magnetic field
${\bf B}$ along the $z$ direction, where we locate a set of $N_v$
rigid vortices (for most of the results presented here, we have
considered values of $N_v$ ranging from $516$ to $4128$). The
dynamics of each vortex line $i$ at position ${\bf r}_i$ can be
described by an overdamped equation of motion of the form
\begin{equation}
\Gamma d{\bf r}_i/dt = \sum_j {\bf f}_{vv}({\bf r}_i - {\bf r}_j)+
\sum_j {\bf f}_{vp}({\bf r}_i - {\bf r}^p_j)+{\bf f}_L({\bf r}_i),
\label{eq:vf}
\end{equation}
where $\Gamma$ is an effective viscosity for vortex flow. The first
term on the right hand side of this equation follows from the fact
that a pair of vortices interact with each other via a long-range
force ${\bf f}_{vv}({\bf r})=AK_1(|{\bf r}|/\lambda)\hat{r}$, where
$A=\Phi_0^2/(8\pi^2\lambda^3)$, $\Phi_0$ is the quantized flux carried
by the vortices, $\lambda$ is the London penetration length, and $K_1$
is a first order modified Bessel function \cite{degennes}. Distances
are always measured in units of $\lambda$. The second contribution
reflects the attractive interaction forces between vortex lines and
quenched in point defects such as oxygen vacancies or other impurities
in the material. These pinning centers are randomly located at
positions ${\bf r}^p_i$ ($i=1,\ldots,N_p$) within the simulation box,
and exert pinning forces according to a Gaussian potential of the form
$V({\bf r}-{\bf r}^p)=V_0\exp[-({\bf r}-{\bf r}^p)^2/\xi^2]$, whose
amplitude and standard deviation are $V_0$ and $\xi$, the
characteristic coherence length of the superconductor, respectively
(The usual number of pinning centers $N_p$ considered is $4128$, and
we have chosen $\xi=0.2\lambda$, characteristic of low temperature
superconductors such as NbSe or NbMo.). Finally, if an external
current ${\bf J}({\bf r})$ is eventually applied to the sample, it
generates a Lorentz-like force acting on the vortices ${\bf f}_L({\bf
r})=\Phi_0 {\bf J}({\bf r})\times \hat{z}/c$, where $c$ is the speed
of light. These coupled equations of motion (\ref{eq:vf}) are
numerically solved with an adaptive step-size fifth order Runge-Kutta
algorithm, imposing periodic boundary conditions in both directions.

\begin{figure}[t]
\centerline{\psfig{file=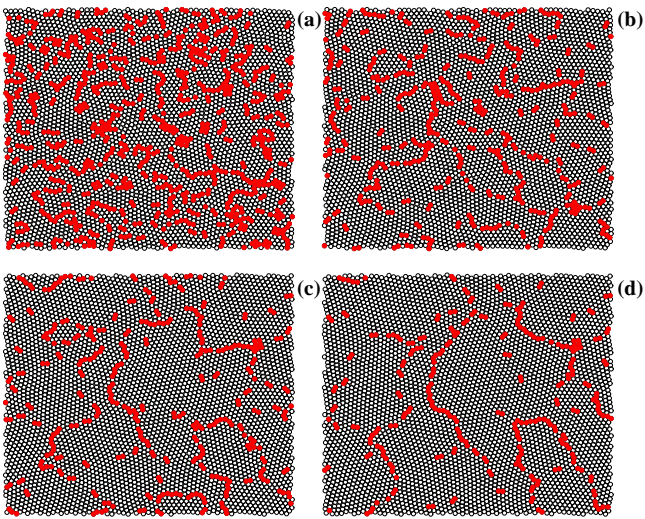,width=8.5cm,clip=!}}
\caption{Relaxation of the topological defect structure from a
simulation of $N_v=4128$ interacting vortices after a sudden field
cooling from a disordered vortex state in a simulation cell of linear
size $L=36\lambda$.  The colored five/seven-fold coordinated vortices
(filled circles) indicate dislocations in the vortex lattice. The
final configuration (snapshot (d)) is completely pinned by
disorder. There one can observe a polycrystalline structure with most
dislocations arranged into grain boundaries.}
\label{snapshots}
\end{figure}

We first consider the relaxation dynamics of the vortex lines in the
absence of driving currents. Moreover, in the present analysis we
completely disregard thermal effects, that is, we mimic the dynamics
of the vortex system after a sudden quench of the superconducting
sample from high temperatures (or equivalently, random vortex
configurations) towards the lower energy states corresponding to zero
temperature. After a transient regime, the dynamics stops due to
disorder. We analyze the resulting spatial configuration of flux lines
by means of Delaunay triangulations. A pair of a five-fold and a
seven-fold neighboring vortices correspond to a dislocation in the
vortex lattice. In the course of the simulations, the number of
five/seven-fold coordinated vortices is the same, indicating that
during the relaxation process no other topological defects such as
disclinations appear to be present in the lattice.

\begin{figure}[t]
\centerline{\psfig{file=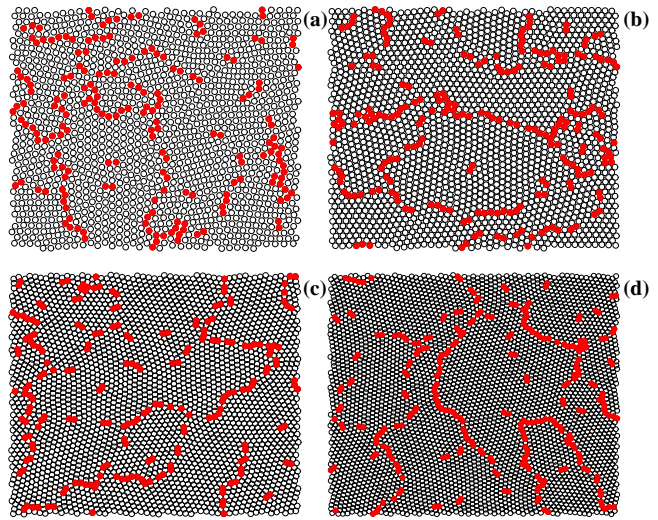,width=8.5cm,clip=!}}
\caption{Pinned vortex structure for different values of the magnetic
field: (a) $N_v=1460$, (b) $N_v=2064$, (c) $N_v=2919$, (d) $N_v=4128$,
after a sudden field cooling from a disordered vortex state in a
simulation cell of linear size $L=36\lambda$.  The colored
five/seven-fold coordinated vortices (filled circles) indicate
dislocations in the vortex lattice. The average grain size in the
resulting polycrystalline structure seems to grow with the intensity of
the average magnetic field inside the cell.}
\label{grainsize}
\end{figure}

In Fig.~\ref{snapshots}, we report a series of snapshots illustrating
that the gradual ordering process involves the arrangement of
dislocations in grain boundaries. The formation of these walls of
dislocations screens out the long range elastic stress and strain
fields otherwise created by dislocations in the lattice and, at the
same time, they render a polycrystalline structure of the vortex
array. This polycrystalline structure evolves in time until the
residual stresses accumulated in the distorted vortex lattice drop
down below the critical value $\sigma_c$. At this point, grain
boundaries get pinned by disorder limiting the average grain size (see
Fig.~\ref{snapshots}(d)). Moreover, the limit grain size $R_g/a$ appears to
increase with magnetic field $B\propto N_v$ (see
Fig.~\ref{grainsize}), in qualitative agreement with experimental
results~\cite{GRI-89} and the theoretical predictions reported in
Ref.~\onlinecite{MOR-04}.

Different experimental, or simulation, protocols will certainly
influence the relaxation dynamics and the resulting metastable
configurations of trapped dislocations and grain
boundaries. Metastability and history-dependent features have been
long recognized in driven vortex lattices~\cite{PAR-97}. We have
considered a field-cooling procedure since most of the Bitter
decoration experiments are performed in a similar manner
\cite{PAR-97,MEN-02,FAS-02}, and can thus be well described by the
current simulations. Nevertheless, other numerical protocols can also
be devised, as for instance the one recently proposed in
Ref.~\onlinecite{CHA-04} to examine the vortex topology across the
so-called peak-effect, that are better suited to reproduce diverse
experimental conditions.

\begin{figure}
\centerline{\psfig{file=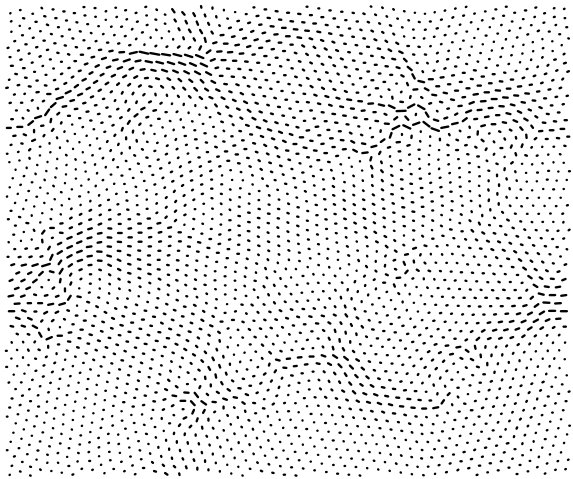,width=7.5cm,clip=!}}
\caption{Vortex trajectories between two pinned configurations
obtained after the application of a small driving current below
the threshold value $J_c(B)$. Small and heterogeneously distributed
displacements of the vortex positions are observed in both the
parallel and perpendicular direction to the applied force $f_L^x$. The
number of vortices in the simulation cell of linear size $L=36\lambda$
is $N_v=2919$. The number of pinning points $N_p=4128$.}
\label{displace}
\end{figure}

Next, we study the behavior of the critical
current $J_c(B)$ for these 2D vortex polycrystals by means of
numerical simulations. An externally applied current may induce the
annealing of the metastable configurations (at least, to a certain
degree that obviously depends on its intensity) present in
Figs. \ref{snapshots} and \ref{grainsize}. This is indeed observed in
our numerical simulations, where we can as well identify the critical
current $J_c$ below which the average motion of the vortex lattice
eventually ceases after a rich initial transient of plastic flow. As
one can observe, for instance, in Fig.~\ref{displace}, a small current
below the threshold value $J_c(B)$ gives rise to non-trivial (i.e. not
just a slight drift along the force direction $f_L^x$) changes of the
displacement field ${\bf u_n}$ of the vortex lattice. The vortex
displacements are heterogeneously distributed and a small component
perpendicular to the force direction can be observed. This in turn,
implies changes of the elastic shear stress distribution responsible
for the Peach-Koehler forces acting on grain boundary dislocations
that, as a consequence, may move and rearrange in response to the new
force field.

\begin{figure}
\centerline{\psfig{file=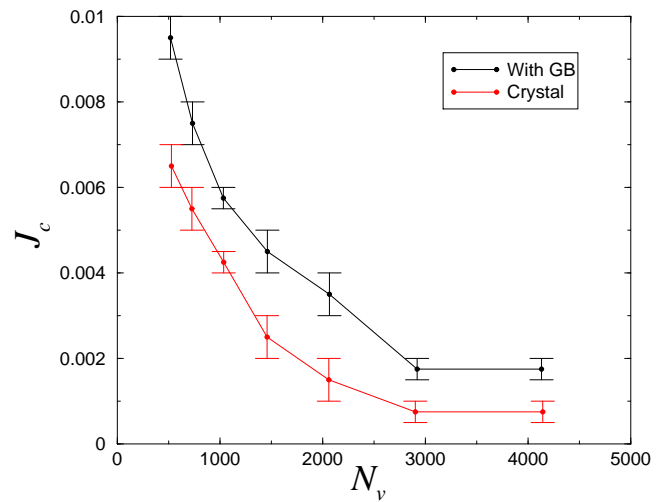,width=8.5cm,clip=!}} \caption{The
critical current $J_c$ as a function of the number of vortices
$N_v$ in the simulation cell. The number of pinning points
$N_p=4128$, the cell size $L=36\lambda$, and the Ginzburg-Landau
parameter considered is $\kappa=5$. The upper line shows the
results obtained starting from initial field-cooled configurations
containing grain boundaries (GB), whereas the lower curve shows
the numerical results obtained from perfect crystalline initial
configurations.} \label{Icr}
\end{figure}

We have determined the dependence of the critical current on the
magnetic field by carrying out simulations for different densities of
vortices in the simulation cell. Moreover, we have compared these
results with those corresponding to a completely different initial
state: a perfect single crystal configuration with similar
densities. Our results are summarized in Fig.~\ref{Icr}. The
qualitative and quantitative differences between the two curves
represented in the figure are due to the presence of grain
boundaries. The presence of these topological defects in the vortex
configuration enhances the critical current needed to give rise to a
steady regime of plastic flux flow, that in this case, appears to be
controlled by grain boundary motion. The plastic deformation of
crystals is usually mediated by the nucleation and motion of
dislocations~\cite{HIR}. Another possible mechanism for plastic flow
is the glide motion of grain boundaries which, as in this case, can be
the most relevant mechanism when the grain sizes are limited and there
is a high fraction of grain boundary atoms. According to our numerical
results, grain boundaries are more efficiently pinned by disorder, in
agreement with the general expectation that grains adjust better
to the disordered landscape than a perfectly ordered lattice.
In both cases, we observe the decrease of $J_c$ with an
increasing density of vortices until this reaches a plateau for the
largest number of vortices considered.

It is also worth noting that we have not considered the renormalization
of either the penetration length $\lambda$ or the coherence length
$\xi$ of the superconductor with the intensity of the magnetic field
$B$. Within a mean-field scenario, these parameters should diverge as
the magnetic field approaches the upper critical field
$B_{c_2}=\Phi_0/2\pi\xi$. An estimation of the reduced field values we
are dealing with in the simulations yields $B/B_{c_2}=2\pi
N_v/(\kappa^2L^2)\sim (0.1-0.8)$. This means that the renormalization
of $\lambda$ and $\xi$ will be especially relevant for the last point
of the curves in Fig.~\ref{Icr}. Recent simulations of similar vortex
lattices in 2D~\cite{CHA-04} show that indeed such a field
renormalization could be responsible for a sudden increase of the
critical current close to the upper critical field $B_{c_2}$.

\begin{figure}
\centerline{\psfig{file=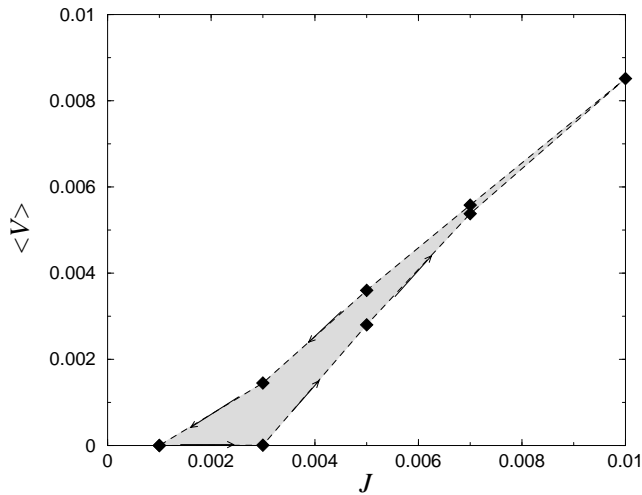,width=8.5cm,clip=!}}
\caption{The steady state average velocity of the vortices as a
function of the applied current $J$. The current is ramped up (and
down) in steps and is kept constant after each step until the
system reaches a steady state. The arrows indicate the direction
of the ramp. The number of vortices is $N_v=2064$, the number of
pinning points $N_p=4128$, the cell size $L=36\lambda$, and the
Ginzburg-Landau parameter considered is $\kappa=5$.}
\label{fig:hyst}
\end{figure}

On an experimental ground, our results match, at least on a
qualitative basis, the behavior exhibited by vortex matter in
critical current measurements at low magnetic fields. As stated
above, grain boundaries are commonly observed in field-cooled (FC)
samples. On the other hand, ordered vortex crystals can be
obtained in zero field cooling (ZFC) experiments, i.e. applying a
magnetic field only after temperature has been lowered to the
expected value
\cite{XIA-01,HEN-96,XIA-00,BAN-99,JOU-00,LIN-01,SAS-00}. The FC
state is usually characterized by a higher critical current and
has been proven to be metastable\cite{HEN-96,BAN-99}. These
aspects result in a peculiar hysteretic behavior commonly observed
in critical current measurements\cite{HEN-96,BAN-99} and $I-V$
characteristics \cite{XIA-01,XIA-00}.  In our numerical analysis,
the evaluation of critical currents in perfect vortex crystals
(lower line in Fig.~\ref{Icr}) fairly mimics phenomenology of ZFC
measurements, while results for the grain boundary model (upper
line in Fig.~\ref{Icr}) can be interpreted as a simulation of FC
response. Hysteresis is in fact reproduced by our simulations when
we start from the polycrystalline state. As shown in
Fig.~\ref{fig:hyst}, when the current $J$ is ramped up vortices
start to move at a current $J_{c1}$, with a velocity that then
increases with the current. If the current is ramped down from the
moving state, vortices get pinned at a lower value of the current
$J_{c2}$ corresponding to the critical current measured for a
perfect crystal upon ramping up the field. Notice the similarity
with the experimental results of Refs.~\cite{XIA-01,XIA-00}. Once
more, we should underline how these results hold only for low
values of the applied field. As the magnetic induction approaches
its critical value, a sudden increase in measured critical
currents is observed in both the ZFC and the FC experimental setup
\cite{HEN-96,BAN-99}.


\section{Grain boundary induced melting}
\label{melting}

The stability of crystalline ordering in a vortex lattice beyond
the well known Bragg Glass regime is still a matter of
investigation. Experimental results suggest that an increase in
temperature above a certain critical value $T_m$ determines the
transition to a {\it liquid} phase \cite{SAF-92,BOC-01,AVR-01},
while the effects of disorder associated with high magnetic fields
are responsible for the insurgence of a {\it glassy} phase
\cite{CUB-93,GAM-98,LIN-01}. A deep theoretical understanding of
such transition phenomena, accounting for their microscopic
origin, has not been achieved yet. Nonetheless it has been shown
that for strong enough disorder the Bragg glass phase is unstable
against dislocation formation \cite{CAR-96,KIE-97,FIS-97}. This
suggests that the melting process could be ruled by topological
defects (as discussed in Ref.~\onlinecite{KIE-00}) in  analogy with two
dimensional theories of crystal melting. Here we discuss the
possibility that, under the effects of fluctuations, dislocations
unbind and rearrange in grain boundaries giving rise to a
polycrystalline structure \cite{CHU-83}. In this framework, the
vortex polycrystal can be seen as an intermediate stage in a
process that ends in the amorphous or liquid phases.

Our purpose is to study the quasiequilibrium properties of such
polycrystalline stage, using the elastic properties of grain
boundaries in a vortex lattice derived in Section II. The main goal of
our analysis is to write the free energy density $f$ of the system as
a function of different lattice arrangements in configuration space. A
minimum in free energy for a polycrystalline configuration in
proximity of the melting line would corroborate the hypothesis of a
grain boundary mediated transition. For our purposes, we parametrize
configuration space in terms of linear grain boundary density $n$,
meaning that a $n\rightarrow 0$ configuration corresponds to an
ordered (grain boundary free) vortex lattice. Our consideration focus
on the thermally induced melting transition and the effects of
impurities are neglected.


We consider arrays of edge dislocations, parallel to the $z$ axis
and arranged in low angle grain boundaries. As in the case of
grain growth, all Burgers vectors are in the $xy$ plane,
corresponding to a columnar grain structure. Following the
aforementioned ideas \cite{CHU-83}, we can introduce the linear
concentration of grain boundaries $n$ and in the low density limit
we can expand the free energy functional (per unit volume) in
powers of $n$ as
\begin{equation}\label{free_energy}
f(n)=\left(\gamma_0+\gamma_T\right)n+ \Gamma\, n^2-M\,\Gamma\, n^3.
\end{equation}
The different coefficients of the expansion are explained in the
following. The linear term is due to the elastic energy of grain
boundaries. The zero temperature contribution $\gamma_0$ is the
elastic energy per unit surface of a flat or smooth grain boundary
that, in the limit of low angle grain boundaries, is given
by\cite{HIR}
\begin{equation}
\gamma_0\simeq\frac{c_{66}b^2}{2\pi D}\ln \frac{e\chi D}{2\pi
b},
\end{equation}
where the $\chi>0$ factor takes into account core interaction
effects. The $\gamma_T$ term, on the other hand, accounts for thermal
fluctuations. Indicating the elastic Hamiltonian in Eq.~(\ref{iso}) as
$\mathcal{H}= \int\frac{d^2k}{(2\pi)^2} \phi({\bf k})X({\bf k})X(-{\bf
k})$ with $\phi({\bf k})=\epsilon | {\bf k}|$, and $\epsilon=\pi b^2
K/2D^2$, the partition function of a thermally perturbed grain boundary
over a surface $S$ is thus
\begin{equation}
\mathcal{Z}=\int\prod_{\bf k}\,du_{\bf k}\,e^{-\beta\phi_{\bf k} u^2_{\bf k}}
\end{equation}
and the corresponding free energy per unit surface
\begin{equation}
\gamma_T=-\frac{1}{\beta S}\ln\mathcal{Z}
\end{equation}
with $\beta=(K_BT)^{-1}$. The above term can be determined explicitly
calculating the logarithm of the partition function as
\begin{eqnarray}
\ln\mathcal{Z}=\frac{S}{2}\left[\frac{1}{Da_z}\ln\left
(\frac{e^\frac{3}{2}}{2\beta\epsilon S \xi^2}\frac{Da_z}{\sqrt{D^2+a_z^2}}
\right) \right.\nonumber\\ \left.+\frac{1}{D^2}
\arctan\frac{D}{a_z}+\frac{1}{a^2_z}\arctan\frac{a_z}{D}\right]
\end{eqnarray}
where we have introduced a short wavelength cutoff $2\pi/a_z$ to
delimit the integration domain along the $z$ axis.

The $\Gamma$ coefficient of the $n^2$ term is proportional to the
energy of a junction between two grain boundaries and details of its
computation will be given in Appendix.

The $n^3$ term captures the case of the intermission of a third grain
boundary in a junction, screening the effect introduced by the $n^2$
contribution. When this is the case, one loses an energy equal to
$\Gamma n^2$ times the probability of such an event. In the low
density limit, this probability is $Mn$, where $M=2\pi/D$ is roughly
the interaction range of a grain boundary \cite{CHU-83}.

\begin{figure}
\centerline{\psfig{file=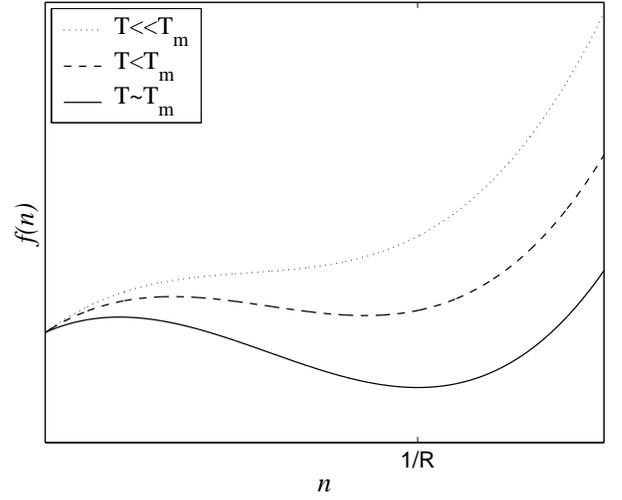,width=8cm,angle=0,clip=!}}
\caption{Free energy density as a function of grain boundary density
close to thermal melting point.}
\label{min1}
\end{figure}

It is convenient to define $\Theta=\ln\mathcal{Z}/S$, so that
the free energy functional in Eq.~(\ref{free_energy}) can be rewritten as
\begin{equation}
f(n)=K_B\Theta\,(T_m-T)\,n+\Gamma\, n^2-M\,\Gamma\, n^3,
\end{equation}
defining a melting temperature as $\displaystyle T_m=\frac{\gamma_0}
{K_B\Theta}$.  As shown in Fig.~\ref{min1}, for values of $T$ close
to $T_m$, $f(n)$ shows a global minimum corresponding to a GB density
$\displaystyle \frac{1}{R}=\frac{1+\sqrt{1+3K_B\Theta
(T_m-T)M/\Gamma}}{3M}$ where $R$ is the average grain size.
As discussed above, this suggests the possibility of a polycrystalline
arrangement before the amorphous phase takes over. As soon as $T$
reaches its melting value $T_m$, the global minimum density becomes of
the order of $D^{-1}$, grains cannot be defined, and the system loses
polycrystalline ordering in favor of a liquid-amorphous phase
characterized by a typical dislocation spacing of order $a$.

\begin{figure}[b]
\epsfig{file=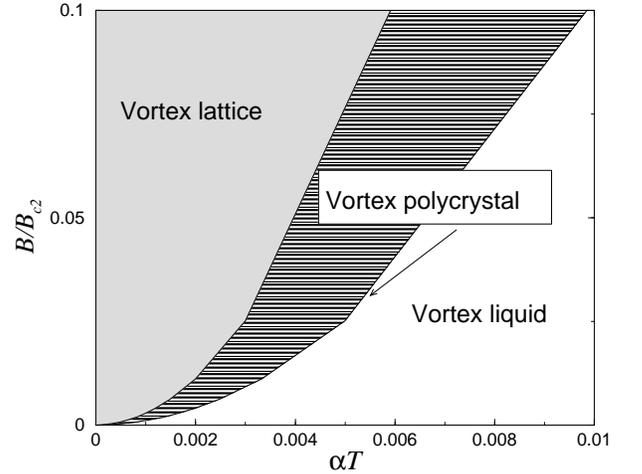,width=8cm,angle=0,clip=!}
\caption{Phase diagram of the vortex ensemble for low values of
the reduced field $B/B_{c_2}$. The temperature is rescaled by the
quantity $\alpha=K_B/(\xi \epsilon_0)$, where $\epsilon_0=(\Phi_0
)^2/(4\pi\lambda)^2$ is an energy per unit length along the
magnetic field direction, i.e. the typical energy for vortex
interactions. The melting line is anticipated by the insurgence of
a polycrystalline ordering.} \label{phase}
\end{figure}

The considerations above allow to draw a phase diagram for the vortex
array at low applied magnetic fields (i.e. when effects of disorder
can be neglected). The resulting plot is shown in Figure
\ref{phase}. The melting line is obtained plotting the above
temperature $T_m$ as a function of the magnetic induction. Here we use
the expression for the local value of $c_{66}$ reported in Ref.~\onlinecite{MIG-00}.
The curve shows reentrant behavior expected for
low fields, due to the exponential decay of the elastic shear modulus
in the $B/B_{c_2}\rightarrow 0$ limit.  The line delimiting lattice
and polycrystal phases, instead, is obtained imposing that the free
energy minimum shown in Fig.~\ref{min1} is a global minimum. In the
presence of disorder, we obviously expect modifications of this
schematic phase diagram. Nevertheless, for weak enough disorder the
main features should remain valid.

\section{Summary and outlook}
\label{conclusions}

In this paper we have investigated the properties of grain boundaries
in a vortex polycrystalline phase. A vortex polycrystal is
experimentally observed in field cooling experiments when grain growth
is arrested by disorder
\cite{GRI-94,GRI-89,MAR-97,MAR-98,PAR-97,FAS-02}, and could also arise
close to the melting line as an intermediate stage between the vortex
lattice and the vortex liquid or glass \cite{DAS-04,CHA-04}.  In both
cases, the dynamics of the system can be studied analyzing grain
boundaries, which play a similar role to that of domain walls in
ferromagnetic systems.  Grain boundaries can be seen as elastic
manifolds whose non-local surface tension can be obtained from the
elastic description of the vortex lattice. Deformations are due to
thermal fluctuations or to random stresses induced by vortex lattice
deformations \cite{KIE-00}.  Once the main ingredients
(i.e. elasticity and disorder) have been properly described, grain
boundaries can be studied with standard scaling methods, used in the
past for various systems from flux lines to ferromagnetic domain
walls. In particular, we have studied disorder induced roughening,
depinning under an applied stress and creep. These results are
important to quantify arrested grain growth due to disorder and can be
used to estimate the grain size in field cooling experiments
\cite{MOR-04}.

An important question concerns the relevance of a polycrystalline
vortex structure for the transport properties of a superconductor
\cite{PAR-97}.  We have shown by numerical simulations that the
critical current of a vortex polycrystal is systematically higher than
the one observed in the corresponding single crystal case. This result
reflects the fact that a polycrystal is pinned more effectively than a
single crystal.

Finally, we have extended the theory of grain boundary induced melting
\cite{CHU-83} to vortex lattices. We have written a free energy as a
function of the grain boundary density considering the contributions
due to thermal fluctuations, elastic deformations, and junction
formation.  We find that the ordered crystal melts into a liquid
passing through an intermediate polycrystalline phase in agreement
with recent numerical results \cite{DAS-04,CHA-04}. We have drawn a
schematic phase diagram as a function of temperature and magnetic
field which, however, can only be considered as a first rough
approximation.  We have not taken into account the effect of disorder,
which is believed to be responsible for a field induced transition to
an amorphous vortex glass. In addition, we have neglected the effect
of isolated dislocations and their interactions with grain boundaries.
Therefore at this stage the present theory should be seen mainly as a
framework for a general physical mechanism \cite{MEN-01},
supported by simulations
\cite{DAS-04,CHA-04} and by some experiments \cite{DIV-04}, for
vortex lattice melting.

\section*{Acknowledgements}
We thank M. Zaiser for discussions and suggestions. This work has
been partially supported by DGES of the Spanish government, Grant
No. FIS2004-05923-CO2-02. M.C.M. acknowledges financial support
from the Ministerio de Educación y Ciencia (Spain), and from the
Departament d'Universitats, Recerca i Societat de la Informació,
Generalitat de Catalunya (Spain).

\appendix{Appendix}
\section*{Estimate of the energy associated to junction formation}
The presence of a $n^2$ term in the free energy functional
[\ref{free_energy}] was first suggested by Chui \cite{CHU-83}, in
order to take into account grain boundary crossing in the framework of
a crystal melting theory. Such a crossing energy consisted of a
thermal contribution due to coupling between fluctuations of
dislocations of crossing grain boundaries.  Nonetheless, Bitter
decoration experiments show that in vortex polycrystals, grain
boundaries primarily rearrange forming junctions, instead of simply
crossing.  The formation of such junctions determines variations in
the overall free energy of the system due to two different
contributions, a zero temperature junction elastic energy and a
thermal part related to fluctuations. In the following, we will
address to these contributions respectively as $\Gamma_0$ and
$\Gamma_T$, being $\Gamma=\Gamma_0+\Gamma_T$.

\subsection{Zero temperature energy}

We assume that because of the short range nature of a grain boundary
stress field, grain boundary interactions are screened for long
distances and we show that forming a junction leads to a
zero-temperature elastic energy gain $\Gamma_0\neq 0$.

The idea is to focus on what happens when two grain boundaries come so
close that they can form a junction. Let us consider a first grain
boundary, e.g. directed along ${\bf j}$ with Burgers vectors ${\bf
b}_n$ such that, ${\bf b}_n\cdot {\bf i}=b_n$, and a single
dislocation, belonging to the other grain boundary, whose Burgers
vector is ${\bf b}'\cdot {\bf i}=-b'\cos\varphi$, being $\varphi$ the
junction angle.

Since grain boundary interactions are short-ranged, we expect
misorientations effects to make no difference in the energy
computation until dislocations come close to a distance that we will
call $\bar{s}$.  If, on a distance $\bar{s}$, the interaction energy
for $\varphi=0$ is lower than for $\varphi\neq 0$, there in no reason
for the system to make a junction. Otherwise, if there is an energy
gain, grain boundaries are likely to join.
\begin{widetext}
Considering the general expression for dislocation interactions
\begin{equation}\label{V}
V=-\,\frac{K}{2\pi}\left[\ln \frac{e\alpha\left|{\bf r}-{\bf
      r'}\right|}{b}\,{\bf b}\cdot{\bf b'}\, -\,\frac{{\bf
      b'}\cdot({\bf r}-{\bf r'})\,{\bf b}\cdot({\bf r}-{\bf
      r'})}{\left| {\bf r}-{\bf r'}\right|^2}\right],
\end{equation}
where ${\bf r}$ and ${\bf r}'$ are the positions of interacting
dislocations, the energy (per unit length) of our system (GB and
rotated dislocation) is
\begin{equation}
E_s(\varphi)=\frac{Kb^2}{4\pi}\left[\left(\sum_{n=-\infty}^{+\infty}\ln\frac{e^2\alpha^2(s^2+n^2D^2)}{b^2}
  -2\sum_{n=-\infty}^{+\infty}\frac{s^2}{s^2+n^2D^2}\right)\cos\varphi
  -\left(\sum_{n=-\infty}^{+\infty}\frac{snD}{s^2+n^2D^2}\right)\sin\varphi\right].
\end{equation}
where $s$ is the distance between the rotated dislocation and the
grain boundary. Moreover, after summing the series,
\begin{equation}
E_s(\varphi)=\frac{Kb^2}{2\pi}\left[\ln\left(\frac{e\alpha D}{\pi b}\sinh\frac{\pi s}{D}\right)
-\frac{\pi s}{D}\coth\frac{\pi s}{D}\right]\cos\varphi.
\end{equation}
\end{widetext}
Assuming that we have $M$ dislocations within the range of $\bar{s}$,
the energy gain due to a junction will be
\begin{equation}
\Gamma_0=\sum_{m=1}^{M}E_{s_m}(\varphi)-ME_{\bar{s}}(0)<0
\end{equation}
Since the stress field generated by a grain boundary is exponentially
suppressed beyond a distance of the same order of the dislocation
spacing, we can give a rough estimate of the sum taking $M=1$ and
$s_1=D$, i.e.
\begin{equation}
\Gamma_0\simeq-\frac{Kb^2}{2\pi}(1-\cos\varphi)\ln\left(\frac{e\alpha
D}{2\pi b}\right).
\end{equation}

\subsection{Thermal fluctuations}
The $\Gamma_T$ contribution, due to the coupling between fluctuations
of dislocations belonging to different grain boundaries in a junction
can also be estimated following Ref.~\onlinecite{CHU-83}. After
performing the thermal average of the interaction potential (\ref{V}),
calculated on the cylinder of radius $|r|<(2/\sqrt{3})D/\pi$ and
taking the short range logarithmic part of $V$
\begin{equation}
\Gamma_T=\mathcal{N}\int dz\int_A dr\,rV\,e^{-\beta V},
\end{equation}
being $\mathcal{N}$ a normalization constant.
Evaluating the integral for $T\simeq T_m$ leads to
\begin{equation}
\Gamma_T\simeq-\frac{Kb^2}{2\pi}\,\cos\varphi\,\ln^2\left(\frac{D}{b}\frac{2}{\pi\sqrt{3}}\right),
\end{equation}
where $\varphi$ is the average junction angle. In the estimate of
the lattice-polycrystal crossover we have assumed
$\varphi\simeq\pi/3$, as it is often observed in decoration
experiments.

\end{document}